\documentclass[aip,apl,twocolumn,groupedaddress]{revtex4}

\usepackage{bm}
\usepackage{graphicx}

\begin{document}

\title{Spin- and valley- coupled electronic states in monolayer WSe$_2$ on bilayer graphene}
\author{K. Sugawara,$^1$ T. Sato,$^2$ Y. Tanaka,$^2$ S. Souma,$^1$ and T. Takahashi$^{1,2}$}
\affiliation{$^1$WPI Research Center, Advanced Institute for Materials Research, 
Tohoku University, Sendai 980-8577, Japan}
\affiliation{$^2$Department of Physics, Tohoku University, Sendai 980-8578, Japan}

\date{\today}

\begin{abstract}	
We have fabricated a high-quality monolayer WSe$_2$ film on bilayer graphene by epitaxial growth, and revealed the electronic states by spin- and angle-resolved photoemission spectroscopy.  We observed a direct energy gap at the Brillounin-zone corner in contrast to the indirect nature of gap in bulk WSe$_2$, which is attributed to the lack of interlayer interaction and the breaking of space-inversion symmetry in monolayer film.  A giant spin splitting of $\sim$0.5 eV, which is the largest among known monolayer transition-metal dichalcogenides, is observed in the energy band around the zone corner.  The present results suggest a high potential applicability of WSe$_2$ to develop advanced devices based with the coupling of spin- and valley-degrees of freedom.

\end{abstract}

\maketitle
Transition-metal dichalchogenides (TMDs) MX$_2$ (M = transition metal, X = chalcogen) have been a target of intensive studies because they show a variety of physical properties such as superconductivity and charge density wave.\cite{Chhowalla} Recently, monolayer MX$_2$ (the thinnest limit of multilayer 2$H$-MX$_2$) has attracted special attention since it provides a platform to realize advanced electronic devices utilizing the spin- and valley-degrees of freedom.\cite{Xiao} Bulk (multilayer) 2$H$-MX$_2$ (M = Mo and W; X = S, Se, and Te) is an ordinary semiconductor with an indirect band gap between the valence-band maximum (VBM) at the $\Gamma$ point and the conduction-band minimum (CBM) at an intermediate point between $\Gamma$ or K(H) points.\cite{Zhu, Mattheiss, Coehoorn, Bocker} On the other hand, it has been theoretically predicted that monolayer MX$_2$ with two-dimensional honeycomb-like network [Fig. 1(a)] would possess a direct band gap between the VBM and the CBM at the K (K') point due to the lack of interlayer interaction and the A/B-site sublattice asymmetry.\cite{Xiao} The strong spin-orbit interaction of $d$ orbital and the broken space-inversion symmetry [{\it E}({\it k}, $\uparrow$) = {\it E}({\it -k}, $\uparrow$)] lift the degeneracy of the valence band around the K and K' points. These split bands have an out-of-plane spin-polarization vector directing oppositely to each other, satisfying the requirement from the time-reversal symmetry [{\it E}({\it k}, $\uparrow$) = {\it E}({\it -k}, $\downarrow$)] \cite{Xiao, Zhu}. Moreover, owing to the valley degree of freedom, electrons at the K (K') point are robust against the phonon scattering connecting the K and K' points.\cite{Xiao} It has been predicted that such spin-split bands host anomalous quantum phenomena like spin-valley-filter effect, anomalous quantum Hall effect,\cite{Xiao} and magnetic-field-controlled spin current,\cite{Sun} owing essentially to the ``real-spin'' nature of valley-coupled electronic states in monolayer MX$_2$ \cite{Xiao} in contrast to the ``pseudo-spin'' nature in graphene.\cite{Pesin} It is thus of great importance to experimentally establish the spin-dependent band structure of MX$_2$ to understand the origin of exotic physical properties and design electronic devices based on the band-structure engineering.

Recently, it has been reported that monolayer MoS$_2$ and WSe$_2$ obtained by exfoliating a bulk crystal have a direct band gap at the K (K') point while no clear band splitting was observed.\cite{Jin, Yeh} In contrast, monolayer MoSe$_2$ grown epitaxially on bilayer graphene \cite{Zhang} and monolayer MoS$_2$ on HOPG \cite{Alidoust} exhibit the band splitting of $\sim$180 meV.  There is no consensus on whether the energy bands at the K (K') point in bulk WSe$_2$ are spin-polarized or not \cite{Iwasa, Riley}. These facts have left an experimental ambiguity regarding the spin-split/polarized nature of the energy bands and its relationship to the exotic physical properties in MX$_2$ \cite{Zeng, Mak}. In this context, monolayer WSe$_2$ is a suitable candidate to access such an issue since WSe$_2$ is expected to have a large spin-orbit coupling due to the heavy atomic mass and hence the spin splitting should be large enough to be experimentally detected.\cite{Zhu, Riley, Iwasa}  It is also noted that the large spin splitting would be a great advantage to realize more effective electronic devices. While angle-resolved photoemission spectroscopy (ARPES) has been applied to both an exfoliated monolayer film and a bulk crystal of WSe$_2$ \cite{Riley, Iwasa, Yeh}, the electronic structure of an epitaxially grown monolayer WSe$_2$ film remains elusive. It is thus of particular importance to elucidate the basic electronic states of eptaxially grown monolayer WSe$_2$ to develop electronic devices based with monolayer MX$_2$. 

In this paper, we report an ARPES study on a monolayer WSe$_2$ film epitaxially grown on bilayer graphene on SiC(0001).  We observed a direct band gap at the K (K') point in the monolayer film, showing the transition of the band gap nature from ``indirect'' to ``direct'' upon reducing the thickness of crystal. We also revealed a giant band splitting of $\sim$0.5 eV at the K (K') point, and confirmed the spin-split nature by the spin-resolved ARPES measurement. We discuss implications of the present results in relation to the physical properties and electronic-structure studies of MX$_2$.


\begin{figure}
\includegraphics[width=6.1in,bb=0 0 490 320]{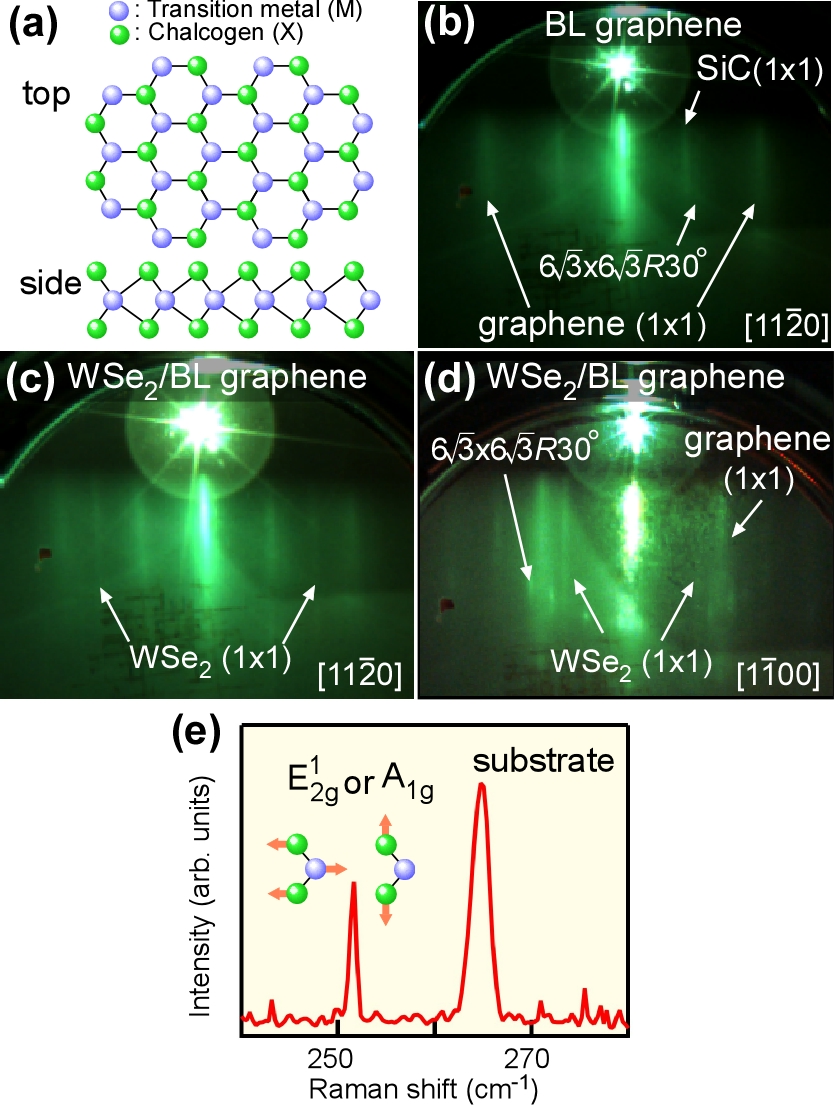}
\caption{(Color online)  (a) Crystal structure of monolayer MX$_2$. (b-d), RHEED patterns of (b) bilayer (BL) graphene and (c-d) monolayer WSe$_2$ on BL graphene. The RHEED patterns in (b) and (c) are obtained along the [1120] azimuth direction (vertical direction) of the SiC substrate while that in (d) was obtained along the [1100] direction. (e) Raman spectrum of monolayer WSe$_2$ on BL graphene.  Inset shows the schematic illustration of the E$^{1}$$_{2g}$ and A$_{1g}$ modes.}
\end{figure}

A monolayer WSe$_2$ film was grown on bilayer graphene by the molecular-beam-epitaxy (MBE) method in an ultrahigh vacuum of 3$\times$10$^{-10}$ Torr.  Bilayer graphene was prepared by annealing an n-type Si rich 6H-SiC(0001) single crystal \cite{JPSJSuga} by resistive heating at 1100 $^\circ$C in a high vacuum better than 1$\times$10$^{-9}$ Torr. Monolayer WSe$_2$ was grown by evaporating tungsten metals on a bilayer-graphene/SiC substrate in selenium atmosphere. The film thickness was calibrated by the deposition rate of tungsten and selenium atoms monitored with a quartz oscillator. The substrate was kept at $\sim$380 $^\circ$C during the sample growth.  The as-grown film was annealed at $\sim$400 $^\circ$C for 30 min, and transferred to the ARPES-measurement chamber without breaking vacuum. The growth process was monitored by the reflection high-energy electron diffraction (RHEED). The characterization of samples was performed by Raman spectroscopy (Horiba LabRAM HR spectrometer). Spin-unresolved ARPES measurements were carried out using a MBS-A1 electron-energy analyzer with a high-flux helium discharge lamp and a toroidal grating monochromator at Tohoku University and a VG-Scienta SES2002 electron-energy analyzer with synchrotron radiation at the beamline BL28A at Photon Factory (KEK). The He I$\alpha$ resonance line ($h\nu$ = 21.218 eV) and the circularly polarized light of $h\nu$ = 60 eV were used to excite photoelectrons. The energy and angular resolutions were set at 16-30 meV and 0.2$^\circ$, respectively. Spin-resolved ARPES measurements were performed with a MBS-A1 electron-energy analyzer with a mini-Mott detector at Tohoku University.\cite{SoumaRSI}  The energy resolution was set at 40 meV.  The sample was kept at 30 K in an ultrahigh vacuum better than 1.0$\times$10$^{-10}$ Torr during ARPES measurements. The Fermi level ($E$$_F$) of sample was referenced to that of a gold film deposited onto the sample substrate. 

Figure 1(b) shows the RHEED pattern of pristine bilayer graphene grown on SiC(0001). We clearly observe the 1$\times$1 and 6$\sqrt{3}$$\times$6$\sqrt{3}$ $\it R$30$^\circ$ spots corresponding to bilayer graphene and buffer layer beneath graphene, respectively.\cite{Zhang} Upon evaporation of tungsten metals on bilayer-graphene in selenium atmosphere, the RHEED pattern exhibits an additional 1$\times$1 streak pattern accompanied with the reduction of the 6$\sqrt{3}$$\times$6$\sqrt{3}$ $\it R$30$^\circ$ spot [Figs. 1(c) and (d)].  As seen in Figs. 1(c) and (d), the observed diffraction patterns strongly depend on the SiC azimuth direction with respect to the incident electron beam. This suggests that a homogeneously ordered monolayer WSe$_2$ film is grown on bilayer graphene. As shown in Fig. 1(e), Raman spectroscopy measurement of this sample has revealed two sharp peaks at 266 and 251 cm$^{-1}$, which are assigned to the substrate \cite{Zhang, Ugeda} and the in-plane or the out-of-plane   mode of monolayer WSe$_2$ \cite{Tongay, Zheng, Yamamoto}, respectively. All these results indicate that a high quality monolayer WSe$_2$ film is fabricated on bilayer graphene.

\begin{figure*}
\includegraphics[width=7.5in, bb=0 0 490 320]{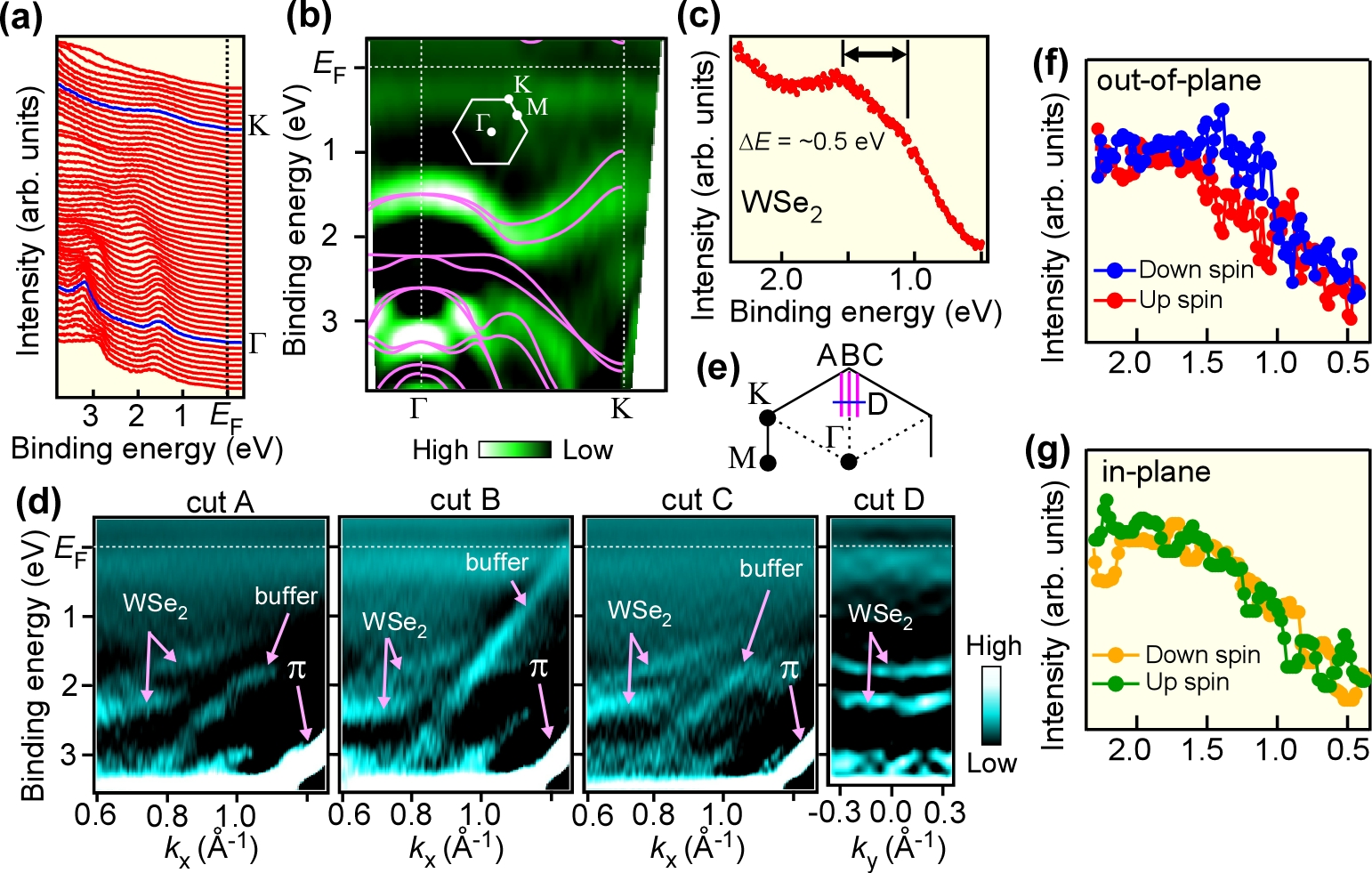}
\caption{(Color online)  (a) ARPES spectra of monolayer WSe$_2$ measured along $\Gamma$K cut with He I$\alpha$ line ($h\nu$ = 21.218 eV) at $T$ = 30 K.  (b) Experimental band structure of WSe$_2$ obtained by plotting the second-derivative intensity of ARPES spectrum as a function of wave vector and binding energy. Solid curves show the first-principles band-structure calculations incorporating the spin-orbit interaction, which were reproduced with a permission from Z. Y. Zhu, Y. C. Cheng, and U. Schwingenschlögl, Phys. Rev. B, 84, 153402 (2011). Copyright 2011 American Physical Society. (c) ARPES spectrum at K point for monolayer WSe$_2$. (d) Experimental band structures derived by plotting the second-derivative intensity of ARPES spectrum as a function of wave vector and binding energy for four representative cuts near K point shown in (e).  (e) Brillouin zone of monolayer WSe$_2$ and the momentum cuts (dashed lines A-D) where the ARPES spectra in (d) were measured. (f) (g) Spin-resolved ARPES spectra of monolayer WSe$_2$ at $T$ = 300 K for the out-of-plane and in-plane spin components, respectively.}
\end{figure*}

Figure 2(a) displays ARPES spectra of monolayer WSe$_2$ on bilayer graphene measured at 30 K along the Γ-K cut with the He I$\alpha$ line ($h\nu$ = 21.218 eV). We clearly observe several dispersive bands, for example, a holelike band which has a top of dispersion at $\sim$1.5 eV at the $\Gamma$ point, and another holelike bands around the $\Gamma$ point at the binding energy higher than 2.5 eV.  The observed systematic dispersive features of energy bands reflect the high single-crystalline nature of the epitaxial WSe$_2$ film. To see more clearly the dispersive features of bands, we have mapped out the band structure by taking the second derivatives of ARPES spectra and plotting the intensity as a function of wave vector and binding energy. As shown in Fig. 2(b), the holelike band with the top of dispersion at 1.5 eV at the $\Gamma$ point shows a characteristic dispersion from the $\Gamma$ to the K points, in good agreement with the first-principles band-structure calculations incorporating the spin-orbit interaction.\cite{Zhu} It is noted that the top of this band at the K point (1-1.5 eV) is closer to $E$$_F$ in comparison with that at the $\Gamma$ point ($\sim$1.5 eV), indicating that the VBM is located at the K point. Since the CBM is likely located also at the K point, the present result strongly suggests that monolayer WSe$_2$ is a direct-gap semiconductor with the valley degree of freedom at K or K' point, forming a massive Dirac cone.\cite{Jin, Yeh, Zhang, Alidoust} This is in sharp contrast to the ordinary indirect-semiconductor nature of bulk WSe$_2$,\cite{Coehoorn} suggesting that the quantum confinement ($i.e.$ the lack of interlayer interaction) and the breaking of space-inversion symmetry convert the nature of band gap from ``indirect'' to ``direct''.  It is noted that the energy difference of the top of the holelike bands between the Γ and K points is ~0.6 eV in monolayer WSe$_2$ while it is 0.1 – 0.4 eV in MoS$_2$, MoSe$_2$, and WS$_2$.\cite{Jin, Yeh}  This suggests that the band gap at the K (K') point in monolayer WSe$_2$ on bilayer graphene has the strongest ``direct-gap'' nature among known monolayer transition-metal dichalcogenides.  This may be related to the strong spin-orbit interaction in WSe$_2$.

The next important question is whether the observed band shows a spin splitting at the K point. As shown in Fig. 2(c), the ARPES spectrum of monolayer WSe$_2$ at the K point measured with the He I resonance line (21.218 eV) shows a peak structure at $\sim$1.5 eV accompanied with a shoulder-like feature at $\sim$1 eV, suggesting the existence of band splitting of $\sim$0.5 eV at the K point. This splitting size is in good agreement with the exciton splitting ($\sim$0.4 eV) in the photoluminescence measurement.\cite{Zheng} To see more clearly the band splitting, we have mapped out the band structure in two-dimensional momentum space around the K point by using synchrotron light of 60 eV which has much smaller beam spot ($i.e.$ higher spatial resolution) and thus more suitable to visualize the band splitting.  A series of experimentally derived band structures measured along four representative cuts shown in Fig. 2(e) signifies a pair of split bands in WSe$_2$, together with the buffer-layer and $\pi$ bands originating from bilayer graphene / SiC.\cite{Zhou} The observed split bands show a similar dispersive feature, as clearly seen in cut D, consistent with the band-splitting scenario.
 
To experimentally establish the spin-split origin of the observed bands, we have performed the spin-resolved APRES measurement with He-I photons.  Figures 2(f) and (g) shows the spin-resolved ARPES results for the out-of-plane and the in-plane spin components, respectively. While the ARPES spectrum for the in-plane spin component [Fig. 2(g)] shows no discernible difference between the up- and down-spin polarization, we observe a small but finite difference in the out-of-plane component as shown in Fig. 2(f), where the down-spin weight is obviously enhanced at the binding energy of 1.0-1.5 eV.  On the other hand, the up-spin component expected to appear at $\sim$1.0 eV is not clearly seen in the spin-resolved ARPES. This may be due to the much weaker spectral intensity of the 1-eV band as compared to that of the 1.5-eV band as seen in Fig. 2(c).  The out-of-plane spin polarization is evaluated to be at most 20$\%$, much smaller than the full spin polarization. This may be explained in terms of mixture of two domains in monolayer crystal which are rotated by 60$^\circ$ from each other. In this case, the ratio of surface area between these two domains is estimated to be 4:6 when we assume that the split band shows a full spin polarization in a single-domain sample.

Now we discuss the spin splitting of WSe$_2$ film in relation to those of other MX$_2$ systems (M=Mo, W: X = S, Se).  We have indentified the spin-lifted valley structure at the K (K') point with the massive Dirac-cone-like feature. The valence band consists of the spin-split bands which switch the spin polarization between the K and K' points to satisfy the requirement from the time-reversal symmetry. The spin splitting observed in WSe$_2$ ($\sim$0.5 eV) is the largest among so-far reported MX$_2$ systems (M=Mo, W: X = S, Se),\cite{Xiao, Zhu, Jin, Zhang, Alidoust} probably due to the large spin-orbit splitting in WSe$_2$.  While a finite spin polarization has been reported in previous spin-resolved ARPES measurements of bulk 2$H$-WSe$_2$ \cite{Riley} and 3$R$-MoS$_2$ \cite{Iwasa}, the valley degree of freedom vanishes due to the indirect gap nature between the VBM at the $\Gamma$ point and the CBM at the K point, and as a consequence the coupling between the real-spin and the valley degrees of freedom does not occur in these indirect-gap semiconductors. In contrast, monolayer WSe$_2$ has a direct gap with a giant spin splitting ($\sim$0.5 eV) much larger than the energy scale of room temperature, which would suppress the spin-valley relaxation caused by the forbidden spin-dependent backscattering between the K and K' points.\cite{Xiao} Thus, monolayer WSe$_2$ has a high potential applicability for realistic spintronic devices operating at room temperature.\cite{Datta, Sun}

Finally, we discuss the implication of present findings in relation to optoelectronics devices. Recently it has been reported that photoluminescence at low temperature in monolayer MoS$_2$ strongly depends on the polarization of circularly polarized light,\cite{Zeng, Mak} owing to the strong coupling between the optical selection rule and the spin-valley coupling.\cite{Xiao} The present ARPES result also suggests that the optical pumping to control the valley polarization would work more effectively at room temperature in WSe$_2$ than in MoS$_2$ because of the large spin splitting. Moreover,  monolayer WSe$_2$ fabricated on graphene has an additional advantage for realizing highly efficient photo-responsive memory devices \cite{Roy} by utilizing the large spin splitting in WSe$_2$ and the transparent nature of graphene.

In conclusion, we reported the spin-resolved ARPES study of monolayer WSe$_2$ grown on bilayer graphene. We have revealed a direct band gap at the K point together with the giant spin splitting ($\sim$0.5 eV) in the energy band, indicating the conversion of the band-gap nature from ``indirect'' to ``direct'' upon reducing the thickness of crystal. The next important step is to develop advanced spintronic devices operating at room temperature based with monolayer WSe$_2$ by utilizing the coupling of spin- and valley-degrees of freedom.

We thank K. Yamada, K. Suzuki, H. Kimizuka, and H. Kumigashira for their help in ARPES experiments.  This work was supported by the JSPS (KAKENHI 23224010 and 24740216), the MEXT (Grant-in-Aid for Scientific Research on Innovative Areas ``Science of Atomic Layers'' 25107003), KEK-PF (Proposal No. 2012S2-001), and World Premier International Research Center, Advanced Institute for Materials Research.

\end{document}